\documentclass[sigconf]{acmart}
\usepackage[super]{nth}
\AtBeginDocument{%
  \providecommand\BibTeX{{%
    \normalfont B\kern-0.5em{\scshape i\kern-0.25em b}\kern-0.8em\TeX}}}

\copyrightyear{2020} 
\acmYear{2020} 
\setcopyright{acmcopyright}\acmConference[DocEng '20]{ACM Symposium on Document Engineering 2020}{September 29-October 2, 2020}{Virtual Event, CA, USA}
\acmBooktitle{ACM Symposium on Document Engineering 2020 (DocEng '20), September 29-October 2, 2020, Virtual Event, CA, USA}
\acmPrice{15.00}
\acmDOI{10.1145/3395027.3419591}
\acmISBN{978-1-4503-8000-3/20/09}

\begin{document}

\title{COVID-19 Kaggle Literature Organization}

\author{Maksim Ekin Eren}
\authornote{Both authors contributed equally to the paper}
\author{Nick Solovyev}
\authornotemark[1]
\affiliation{%
  \institution{Univ. of Maryland, Baltimore County}
}
\email{meren1@umbc.edu}
\email{sonic1@umbc.edu}

\author{Edward Raff}
\affiliation{%
  \institution{Booz Allen Hamilton}
}
\affiliation{%
  \institution{Univ. of Maryland, Baltimore County}
}
\email{raff\_edward@bah.com}

\author{Charles Nicholas}
\author{Ben Johnson}
\affiliation{%
  \institution{Univ. of Maryland, Baltimore County}
}
\email{nicholas@umbc.edu}
\email{ben.johnson@umbc.edu}


\begin{abstract}
The world has faced the devastating outbreak of Severe Acute Respiratory Syndrome Coronavirus-2 (SARS-CoV-2), or COVID-19, in 2020. Research in the subject matter was fast-tracked to such a point that scientists were struggling to keep up with new findings. 
With this increase in the scientific literature, there arose a need for organizing those documents.
We describe an approach to organize and visualize the scientific literature on or related to COVID-19 using machine learning techniques so that papers on similar topics are grouped together. By doing so, the navigation of topics and related papers is simplified. We implemented this approach using the widely recognized CORD-19 dataset to present a publicly available proof of concept.
\end{abstract}



\maketitle
\section{Introduction}
In 2020, the world began to endure a health crisis of unprecedented scale. The scientific community was quick to respond. By one estimate, more than 23,000 papers related to COVID-19 were published between January \nth{1} and May \nth{13} of 2020. This figure is doubling every 20 days \cite{Brainard2020}. The influx of information left health professionals and the scientific community overwhelmed and unable to quickly find important publications. In this paper, we present a proof of concept tool that organizes documents on or related to COVID-19 through clustering and dimensionality reduction. Clustering the texts and using the results as labels for a 2-dimensional scatter plot groups similar papers together, and separates dissimilar ones. To this extent, we parsed the COVID-19 Open Research Dataset (CORD-19) \cite{Kaggle}, a collection of publicly available COVID-19 literature.  

In response to the COVID-19 pandemic, the Allen Institute for AI \cite{ALLENAI}, the Chan Zuckerberg Initiative (CZI), Georgetown University’s Center for Security and Emerging Technology, Microsoft Research, IBM, and the National Library of Medicine - National Institutes of Health prepared the CORD-19 dataset. CORD-19 consists of scientific literature which embodies over 141,000 scholarly articles related to the Corona family of viruses. The White House Office of Science and Technology Policy put out a call to action from citizen scientists to help better understand the virus through analysis using data science techniques \cite{WHITEHOUSE}. The dataset was publicly hosted on Kaggle\footnote{Kaggle, a subsidiary of Google LLC, is a collaborative data science website. \url{kaggle.com}}. Users of the site quickly responded to the call to action and began experimenting with the data. Our analysis focused on using natural language processing, clustering, and dimensionality reduction methods to produce a structured organization of the literature.    

SciSpacy's \footnote{SciSpacy refers to SpaCy \cite{neumann2019} models for biomedical text processing.} biomedical parser was used to format the texts. After pre-processing,  term frequency–inverse document frequency (tf-idf) \cite{ref1} was used to vectorize the body text of the papers, giving for each document a feature vector $X_1$. A reduced feature vector $X_2$ was then produced for each document by applying Principle Component Analysis (PCA) \cite{Jolliffe1986,Binongo2003} but keeping 95\% variance. The $X_2$ vectors were then clustered with k-means \cite{Lloyd82leastsquares} to produce plot labels.  The $X_1$ vectors were reduced to two dimensions for visualization purposes with t-distributed stochastic neighbor embedding (t-SNE) \cite{vanDerMaaten2008}. The output of these algorithms was displayed in an interactive plot that allowed users to quickly access the original papers and navigate related work \footnote{Interactive plot on GitHub: \url{https://maksimekin.github.io/COVID19-Literature-Clustering/plots/t-sne_covid-19_interactive.html}}.

Our solution was first hosted on the Kaggle web site\footnote{\url{https://www.kaggle.com/maksimeren/covid-19-literature-clustering}}, and we received positive feedback from a surprising range of sources \cite{kratsios2020}. Several users have told us how this tool has helped them in their work and research, including an application to other domains in biology. The {\em Jupyter} notebook\footnote{\url{https://github.com/MaksimEkin/COVID19-Literature-Clustering}} we built continues to be forked, or at least visited, at least ten times a day, on average, over the last several weeks. The rest of this paper details the approach and discusses the results.

\section{Related Work}
Visualization of high-dimensional data has proven to be a common objective across many domains. Plotting a representation of data, reduced to two or three dimensions, can provide valuable insights for data scientists. Most common approaches try to preserve the topology of the data. These algorithms map points in one space to another space in an effort to preserve the latent structure of the data \cite{GoodhillSejnowski1996}. Another common objective of these approaches is to preserve local and global structure in a single map, such that there is a clear distinction in differing instances. Preserving the topology in this way can allow for a comprehensive visualization, making it easier to find similar instances as well as dissimilar ones.

Several visualization and search tools using CORD-19 have already been attempted.  \citet{explorer} uses literature-based discovery (LBD) \cite{biolbd} to create a graph of interconnected topics found in the CORD-19 data. \citet{daniel} uses Latent Dirichlet Allocation (LDA) \cite{lda} to form a metric of similarity between papers. Another topic modeling project used a different data source to analyze the discussions of COVID-19. \citet{Ordun2020} utilized LDA, UMAP \cite{McInnes2018,Nolet2020}, and other approaches to visualize the COVID-19 related posts on Twitter. 

Scientific literature visualization has also been used across broader domains. Paperscape, for example, visualizes 1.7 million scientific research articles \cite{Paperscape}. The papers in this tool are organized by way of references/citations to each other. Such clustering will not capture the content of the documents as it does not delve into the text. In our work, we utilize the body text of each paper to better determine the topic similarity.
In another example \cite{Shakespeare}, Tucker tensor decomposition \cite{KoBa09} is utilized to cluster Shakespeare's plays in an effort to demonstrate how the application of tensors and their decompositions might support cluster analysis of malware specimens. We build upon the idea of clustering similar documents within our work.

\section{Methods}
Our analysis pipeline consists of several steps where we clean our dataset, vectorize it to be used by machine learning algorithms, get labels for the documents, and finally visualize our results. The approach described in this paper took approximately 8 hours to run and was computed on a machine with \textit{Intel(R)} Core(TM) i7-4790K CPU @ 4.00GHz and 16GB DDR3 RAM. Here we present the steps within our pipeline.

\subsection{Data Pre-processing}

Our pre-processing approach begins with a dataframe of \textit{n} papers. We first removed duplicates, papers that only contain an abstract, and papers written in some language other than English. Removing non-English papers was specifically important as they would be clustered separately from the rest of the papers on account of language difference alone. After cleaning the dataset, we obtain a corpus of 49,967 papers. The body text of each instance was then tokenized using Spacy's parser. In the tokenization process we removed function (stop) words, or frequently occurring words carrying only trivial semantic meaning. A few examples of stop words as found in our dataset are "doi", "fig", and "medrxiv". We have also removed punctuation, and capitalization.

The next step was then to vectorize the tokenized texts using Scikit-Learn’s \cite{pedregosa2011scikit} tf-idf, creating feature vector {$X_1$}. Tf-idf converts the string formatted data into a measure of how important each word is to the instance out of the literature as a whole. In tf-idf, \textit{n\_features} were limited to $2^{12}$ to limit memory usage and act as a noise filter.

\begin{figure}[tbh]
  \centering
  \includegraphics[width=1\linewidth]{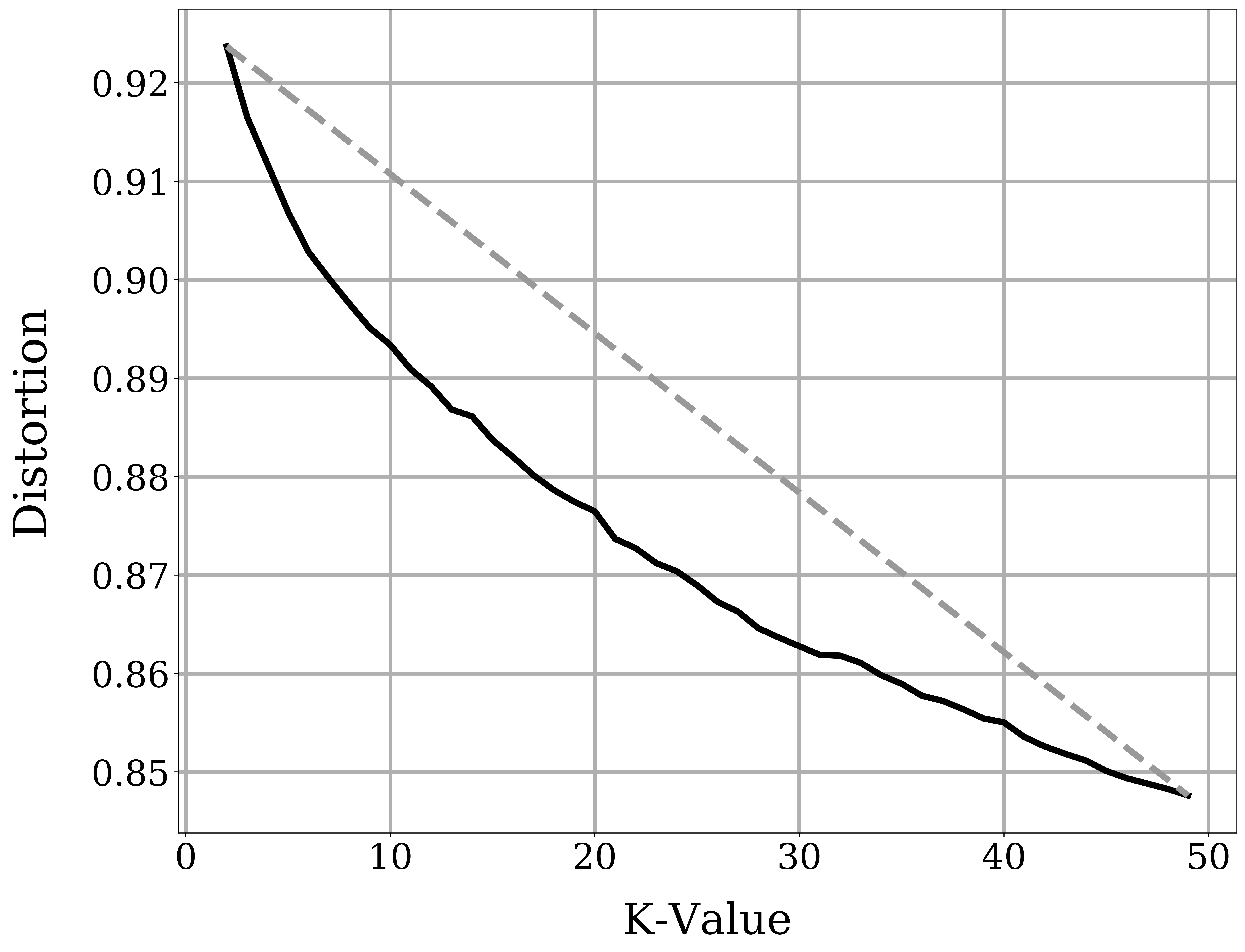}
  \caption{Elbow curve (solid line) showing the distortion score (y-axis) as the number of clusters $k$ (x-axis) changes. The improvement in distortion rapidly decreases until $k=20$. \label{fig:elbow_curve}}
  \Description{}
\end{figure}

After vectorization, PCA was applied to $X_1$ to reduce the dimensions just enough to preserve 95\% of the variance, forming a new embedding $X_2$. PCA finds \textit{x} principal components, each orthogonal to each other. The goal of each principal component is to maximize the variance in the dataset which that component "explains". Once the principal components have been identified, the dataset can be reduced to \textit{d} dimensions by projecting it onto a hyper-plane defined by the first \textit{d} principal components \cite{Aurelien2019}. By keeping a large number of dimensions with PCA, much of the information is preserved, while some noise/outliers are removed from the data.

\subsection{Getting the Labels}

\begin{figure*}[tbh]
  \centering
  \includegraphics[width=1\linewidth]{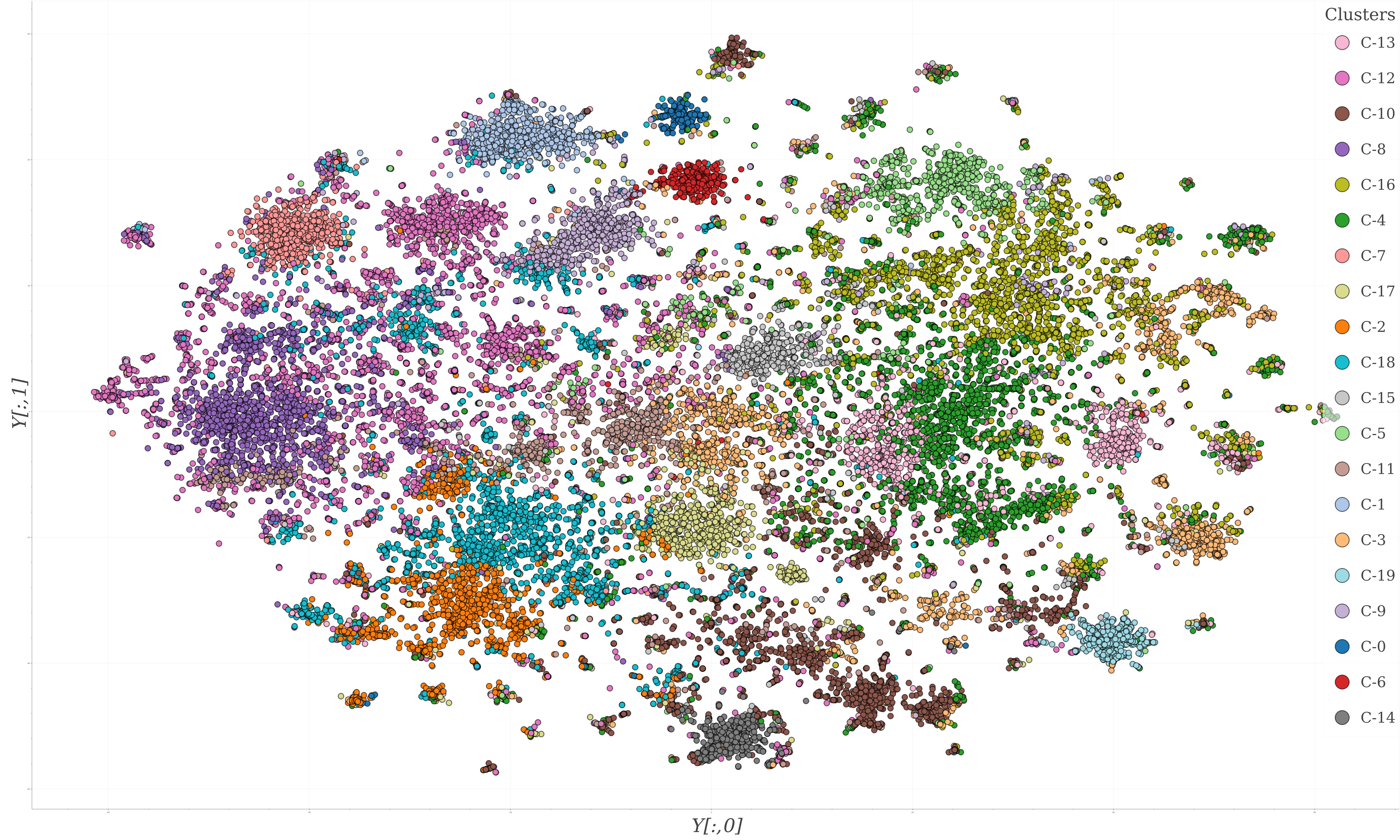}
  \caption{t-SNE is used to visualize COVID-19 related scientific literature. Labels are obtained via k-means clustering of the documents. Papers that explore the same or similar topics are plotted near each other forming clusters. \label{fig:covid_tsne}}
  \Description{}
\end{figure*}

By clustering the literature, we can generate labels for similar papers. When the dataset is reduced to two dimensions, some clusters overlap or are not easily distinguishable within the scatter plot. To reduce this effect, we used k-means, a popular iterative, unsupervised clustering algorithm. First, $k$ instances are randomly selected as initial centroids. An instance is said to belong to cluster $k_i$ if centroid $k_i$ is closer to that instance than any other centroid by the $l2$ norm. The centroids are then updated 
at the end of each pass.
This clustering approach was applied on the feature vectors $X_2$ to get each of the research article $d_i$'s label $y_i$. 

To use k-means, the value of $k$ must be specified, and we determined this value using distortion. Distortion is the sum of the squared distances between every instance vector and its closest centroid. Increasing the number of clusters will reduce the distortion, but this may come at the cost of splitting valid clusters in half. It is common practice to pick a value of $k$ that is on the inflection point of the curve \cite{Aurelien2019}. After the inflection point, improvements in distortion are minimal and decrease with each additional added cluster. As shown in Figure \ref{fig:elbow_curve} this inflection point is approximately $k=20$.

\subsection{Visualization}

We used PCA to perform dimensionality reduction and make the clustering problem easier for k-means, as described above. For visualization purposes we resorted to t-SNE, a popular, non-linear projection algorithm. The t-SNE algorithm generates a probability distribution that represents the interrelationships of the data, and then creates a low-dimensional space that follows this probability distribution. This approach combats the crowding problem \cite{vanDerMaaten2008}. Using t-SNE, the feature matrix $X_1$, consisting of 49,967 instances, is projected down to two dimensions, $X_1$ embedding $Y$, which then can be plotted on the Cartesian plane. Figure \ref{fig:covid_tsne} is a scatter plot of the t-SNE reduced data with k-means results as the labels.

\section{Results}

A common challenge with dimensionality reduction techniques is the verification of results. Often these approaches work well on synthetic datasets, but fall short when applied to real world data. In our study, we tried to verify the success of our work in two ways: manually exploring the clusters and identifying what professionals could impart from our tool.

\subsection{Manual Analysis}
A qualitative check was performed through manual analysis of the t-SNE visualization of CORD-19. In Figure \ref{fig:covid_tsne}, clusters are visually apparent. For example, the red cluster C-6, contains papers on infectious diseases carried by bats. The publications in C-6 are not limited to just coronavirus studies in bats; literature on other viruses like lyssaviruses and filoviruses observed in bat populations is also present within the cluster. We noted with interest that t-SNE grouped papers covering specific strains of virus in close proximity. A sub-cluster within C-6, for example, includes papers dealing with SARS in bats.

We also noted agreement between the clustering labels and the clusters that can be seen on the plot. t-SNE and k-means were applied independently to $X_1$ and $X_2$ respectively. While t-SNE reduced the data to two dimensions, k-means clustered a higher dimensional representation of the data. Despite this, both methods were able to agree on the bounds of many of the more apparent clusters formed by t-SNE. 

However, disagreement between the two approaches is still visible in some sections of the plot. This can be explained by two factors:  
In a visualization problem like this one, the decision boundaries are not clear. The k-means elbow method indicates that using $k=20$ is an optimal solution. Using a higher k-value would undoubtedly result in lower distortion, but this could come at the price of splitting larger clusters in half. As a result, smaller clusters can be detected by t-SNE but might be identified as part of one or several larger clusters by k-means. The other, perhaps more apparent factor, is that k-means was applied in a higher dimensional space, the structure of which cannot be accurately conveyed in a two dimensional plot. 

\subsection{Utilization}

We had the help from Utah State University biomedical engineers Andrew James Walters and Dylan Ellis to further understand the content of the clusters. Using the plot, Walters and Ellis were able to provide insights into the topics and their relationships across different clusters. Niche topics like animal infection and viral reservoirs tended to have tightly grouped clusters. This phenomenon most likely occurred due to a high usage of topic-specific terminology such as the species names. Clusters with multi-disciplinary topics such as epidemiology and virology tended to be either more spread out or shared an overarching theme between different clusters. Using the t-SNE plot, Walters and Ellis also found specific information that is not insignificant. Several new papers described the effectiveness of different drugs and broad spectrum antivirals that could combat COVID-19 symptoms. The plot allowed Walters and Ellis to quickly scan related work and find these publications. 

Our work has also been applied to different domains in biology. Grace Reed, a biology researcher from the University of California, Santa Cruz, uses our techniques to cluster literature on the evolution of tropical reef fish and their response to plastic pollution. Reed's work focuses on exploiting the resulting clusters to identify the topics which will form the basis for future research in the subject matter. 

\section{Conclusion}

We have developed a tool for COVID-19 literature visualization. It is easy to use, performs well qualitatively, and can be applied to other domains. The tool is able to map papers based off of latent interrelationships in the text. Feedback from professionals has indicated it useful to find new and relevant material. 

\bibliographystyle{ACM-Reference-Format}
\bibliography{references.bib}


\end{document}